\documentclass[twocolumn]{aastex631}
\usepackage{xspace}
\usepackage{graphicx}

\usepackage{xcolor}
\usepackage{float}

\usepackage{threeparttable}
\usepackage{longtable}
\usepackage{multirow}

\shorttitle{Atmospheric Escape from the L~98-59 Planets}
\shortauthors{Fromont et al.}

\graphicspath{{./}{figures/}}

\begin{document}

\title{Atmospheric Escape From Three Terrestrial Planets in the L~98-59 System}

\author[0000-0002-3099-0493]{Emeline F. Fromont}
\affiliation{Department of Astronomy, University of Maryland, College Park, MD 20742, USA}
\affiliation{NASA Goddard Space Flight Center, Greenbelt, MD 20771, USA}

%% KEY AUTHORS %%
\author[0000-0003-2086-7712]{John P. Ahlers}
\affiliation{NASA Goddard Space Flight Center, Greenbelt, MD 20771, USA}
\affiliation{University of Maryland, Baltimore County, Baltimore, MD 21250, USA}
\author[0000-0002-8341-0376]{Laura N. R. do Amaral}
\affiliation{Instituto de Ciencias Nucleares, Universidad Nacional Autónoma de México, Cto. Exterior S/N, C.U., Coyoacán, 04510 Ciudad de Mexico, CDMX}
\author[0000-0001-6487-5445]{Rory Barnes}
\affiliation{Department of Astronomy, University of Washington, Seattle, WA 98105, USA}
\author[0000-0002-0388-8004]{Emily A. Gilbert}
\affiliation{Jet Propulsion Laboratory, California Institute of Technology, 4800 Oak Grove Drive, Pasadena, CA 91109, USA}
\author[0000-0003-1309-2904]{Elisa V. Quintana}
\affiliation{NASA Goddard Space Flight Center, Greenbelt, MD 20771, USA}
\author[0000-0002-1046-025X]{Sarah Peacock}
\affiliation{University of Maryland, Baltimore County, Baltimore, MD 21250, USA}
\affiliation{NASA Goddard Space Flight Center, Greenbelt, MD 20771, USA}
\author[0000-0001-7139-2724]{Thomas Barclay}
\affiliation{NASA Goddard Space Flight Center, Greenbelt, MD 20771, USA}
\affiliation{University of Maryland, Baltimore County, Baltimore, MD 21250, USA}
\author[0000-0002-1176-3391]{Allison Youngblood}
\affiliation{NASA Goddard Space Flight Center, Greenbelt, MD 20771, USA}
%

%%%%%%%%%%%%%%%%%%%%%%%%%%%%%%%%%%%%%%%%%%%%%%%%%%%%%%%%%%%%%%%%%%%%%%%%%%%%%%%%

%%%%%%%%%%%%%%%%
% ABSTRACT
%%%%%%%%%%%%%%%%
\begin{abstract}

A critically important process affecting the climate evolution and potential habitability of an exoplanet is atmospheric escape, in which high-energy radiation from a star drives the escape of hydrogen atoms and other light elements from a planet's atmosphere. L~98-59 is a benchmark system for studying such atmospheric processes, with three transiting terrestrial-size planets receiving Venus-like instellations (4--25 S$_\earth$) from their M3 host star. We use the \texttt{VPLanet} model \citep{barnesvpl} to simulate the evolution of the L~98-59 system and the atmospheric escape of its inner three small planets, given different assumed initial water quantities. We find that, regardless of their initial water content, all three planets accumulate significant quantities of oxygen due to efficient water photolysis and hydrogen loss. All three planets also receive enough XUV flux to drive rapid water loss, which considerably affects their developing climates and atmospheres. Even in scenarios of low initial water content, our results suggest that the James Webb Space Telescope (JWST) will \added{be sensitive to observations of}\deleted{likely observe} retained oxygen on the L~98-59 planets in its future scheduled observations, with planets b and c being the most likely targets to possess an extended atmosphere. Our results constrain the atmospheric evolution of these small rocky planets, and they provide context for current and future observations of the L~98-59 system to generalize our understanding of multi-terrestrial planet systems. 
\end{abstract}

%% The AAS Journals now uses Unified Astronomy Thesaurus concepts:
%% https://astrothesaurus.org
%% You will be asked to selected these concepts during the submission process
%% but this old "keyword" functionality is maintained in case authors want
%% to include these concepts in their preprints.
%\keywords{Atmospheric escape, Exoplanet astronomy, Planetary atmospheres}

 %%%%%%%%%%%%%%%%
 % INTRODUCTION
 %%%%%%%%%%%%%%%%
\section{Introduction} \label{sec:intro}

To date, both ground-based radial velocity (RV) surveys and space-based transit surveys have found small \citep[R$<$1.6 R$_\earth$;][]{cloutier2020valley} planets at higher occurrence rates around M dwarfs than for hotter stars  \citep{bonfils2013, dressing2013, dressing2015, hardegree-ullman2019, ment2023}. As of June 2023, NASA's Exoplanet Archive lists 181 confirmed and 116 candidate small planets around M dwarf stars. Additionally, many of these planets orbit nearby, bright stars, making them ideal candidates for atmospheric characterization by JWST \citep{jwst2006}.

While hundreds of multi-planet systems have been discovered, only a limited number are amenable to follow-up observations. Most multi-planet systems have host stars that are too faint or have predicted signal amplitudes that are too small to be useful for follow-up mass and atmospheric measurements.

\begin{figure}[h]
    \centering
    \resizebox{8cm}{!}{\includegraphics{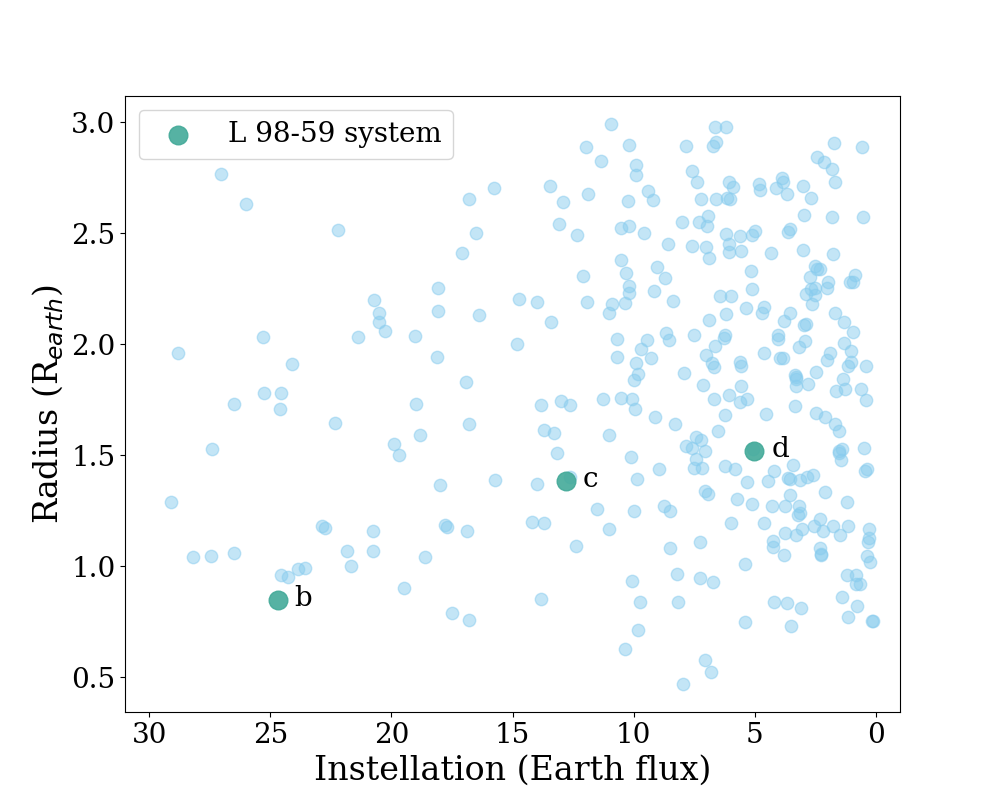}}
    \caption{\added{The radii and instellations of confirmed, transiting, multi-planet systems with planet radii \textless 3 R$\earth$ and instellation \textless 30 Earth fluxes. The L~98-59 planets are labeled in dark green with their corresponding names. Many of these planets are found at lower (\textless 12 Earth flux) instellations, but the L~98-59 planets are spread out over a significant instellation range. The variation in radius and instellation within their same system presents an opportunity to study the evolution of different planets in the same stellar environment.} \deleted{The radii and instellations of select terrestrial planet systems with confirmed mass measurements that have been detected around M dwarfs. Planets within the same system share the same colored marker, and the inner planets of the L~98-59 system are labeled accordingly. Most of these planets occupy a similar instellation range, but L~98-59 b and L~98-59 c are highly-irradiated outliers due to their proximity to their star. The variation in radius and instellation within their same system presents an opportunity to study the evolution of different planets in the same stellar environment.}}
    \label{fig:planet_rad}
\end{figure}

Situated at the border of JWST's continuous viewing zone \citep{demangeon2021}, the L~98-59 system is an excellent benchmark target for JWST observations. As shown in Figure~\ref{fig:planet_rad}, the L~98-59 planets occupy an enticing region of parameter space for follow-up study. The system is composed of four confirmed planets, L~98-59 b, L~98-59 c, L~98-59 d, and L~98-59 e, which orbit around their bright (K = 7.1 mag) and nearby \citep[10.6 pc;][]{kostov2019b} M3 dwarf host star. The inner three small exoplanets, L~98-59 b, L~98-59 c, and L~98-59 d reside interior to the system’s habitable zone (HZ). Our work focuses only on these three planets, as L~98-59 e likely does not transit \citep{demangeon2021}.

% Plotting other small, rocky planet systems around M dwarf:

There is considerable value in modeling the evolution of the L~98-59 system, particularly because of a unique combination of characteristics including the presence of multiple terrestrial planets, their likely exposure to flares, and the system's overall ease of observation. These characteristics will allow us to better understand the interactions of planets around M dwarfs, climate development on Earth-analogs, and potential habitability conditions that arise from such an environment.

\begin{figure}[h]
    \centering
    \resizebox{8cm}{!}{\includegraphics{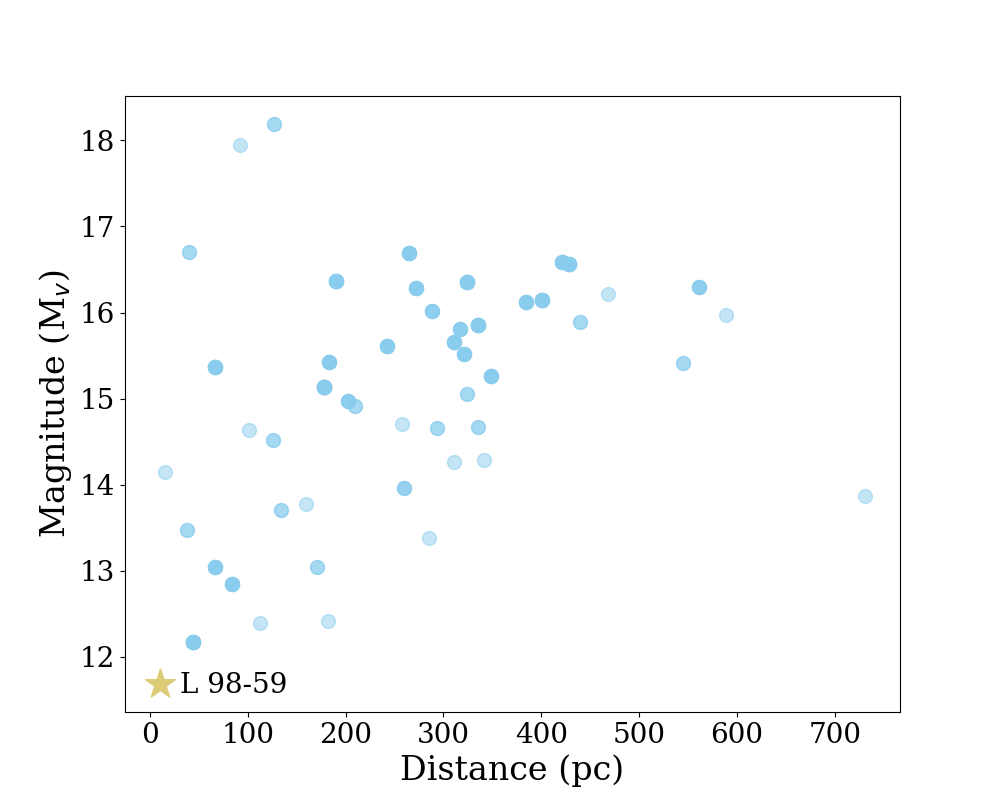}}
    \caption{The distance from the Sun and V-band magnitude of \deleted{select M dwarf stars with multi-planetary systems, with each system labeled correspondingly (colors correspond to}\added{the stars of the planetary systems plotted in} Figure~\ref{fig:planet_rad}. Most of these systems are very faint and/or very distant, making them difficult to observe as atmospheric characterization targets. L~98-59 is both nearby and bright, making it a prime target for JWST observations.}
    \label{fig:star_dist_mag}
\end{figure}

The L~98-59 system presents an excellent opportunity to study the evolution and atmospheric characterization of multiple, small planets that formed in the same stellar environment \citep{greene2016, morley2017, demangeon2021}. L~98-59 is the brightest and nearest M dwarf star system that has at least two measured planetary masses and radii, as seen in Figure \ref{fig:star_dist_mag}. While L~98-59 appeared to be a quiet star when it was first observed by TESS \citep{Ricker2015,kostov2019b}, subsequent observations revealed stellar activity in the form of white-light flares. M dwarf stars produce frequent flares across the electromagnetic spectrum \citep[e.g.][]{Muirhead18,doamaral2022} and remain active for long portions of their lifetimes \citep{west2008, west2015}. This activity likely affects planet atmospheres and must be accounted for in our atmospheric evolution models. Studying the atmospheres of planets orbiting active stars such as L~98-59 will provide us with a greater understanding of the effects of high-energy, or short-wavelength, radiation on the atmospheric retention and evolution of terrestrial planets over time.

% Plotting other well-studied M dwarfs:

% Brief intro to atmospheric escape    
Atmospheric escape is an important process to consider as part of the evolution and potential habitability of exoplanets. Strong X-ray and extreme ultraviolet (XUV; 1-1000 \AA, \citet{ribas2005}) radiation from the host star drives heating and ionization of the upper atmospheres of highly irradiated planets, leading to the escape of gases \citep{luger2015, koskinen2022}. XUV-driven escape is believed to have strongly affected the atmospheric evolution of solar system planets, such as Venus, Earth, and Mars \citep{lammer2008}, and is considered to be a likely culprit for sculpting the observed exoplanet population \citep{owen2020}. Indeed, planets orbiting close to their host stars, like the L~98-59 planets, are particularly vulnerable to atmospheric escape. Transit observations by the Hubble Space Telescope (HST) in the far-UV (FUV) spectrum confirm that many planets close to their host stars lose mass to hydrodynamic escape (e.g., HD 209458b \citep{vidalMadjar2003}, HD 189733b \citep{LecavelierdesEtangs2012}, GJ 436b \citep{Ehrenreich2015}, GJ 3470b \citep{bourrier2018}), which can have important consequences not only for planets' atmospheric evolution, but for their structure and  composition as well \citep{luger2015}.

If the inner three planets of the system contained any volatiles in their initial composition, they should experience atmospheric escape because of their proximity to their star. To study the impact of XUV-driven escape on the L~98-59 planets, we use the modeling tool \texttt{VPLanet} \citep{barnesvpl} to evolve the system over billions of years. By studying the atmospheric loss of this system, we hope to better understand the effects of an M dwarf's stellar environment on the development of terrestrial-size planets, which may impact their promise as potential targets for biosignature detection.

In the following sections, we describe our findings on the atmospheric evolution of the three inner planets in the L~98-59 system. We start by discussing the observational data and system properties of the L~98-59 system in Section \ref{sec:observations}. We then describe our planetary simulation model in Section \ref{sec:model}. We interpret our model's results in Section \ref{sec:results} and discuss their implications for future studies of the system in Section \ref{sec:discuss}. Finally, we conclude with an overview of our results and their importance in the context of a new era of JWST observations in Section \ref{sec:conclu}.

%%%%%%%%%%%%%%%%
 % SYSTEM PROPERTIES AN8D OBSERVATIONS
%%%%%%%%%%%%%%%%
\section{System Observations and Properties} \label{sec:observations}

 % OBSERVATIONS  %%%%%%%%%%%%%%%%
\subsection{Observations} \label{subsec:observations}

% TESS observations
TESS observed L~98-59 (TIC 307210830, TOI 175) \added{for 21 sectors, up to and including Sector 69 \citep{TessTool2020}}\deleted{in Sectors 2, 5, 8, 9, 10, 11, 12, 28, and 29 \citep{demangeon2021}}. \citet{kostov2019b} initially reported the system's discovery, in which they confirmed the presence of the inner three terrestrial planets (b, c, d). Since then, \citet{cloutier2019} and \citet{demangeon2021} have provided planetary masses and eccentricities (Section~\ref{subsec: planet prop}). Although \citet{demangeon2021} has recently confirmed the fourth planet, L~98-59 e, (likely a rocky planet or a water world), we do not consider this planet in our model and analysis because it does not transit. We set our stellar and planetary parameters using the values displayed in Table \ref{tab:initparams}.

% Tom's observations
\added{Partly} owing to L~98-59's relative brightness for a host of small planets, it is an excellent transmission spectroscopy target with HST and JWST \added{(see Section \ref{subsec:discussion_volatiles} for more detail)}. No atmospheric signal was conclusively detected by HST/WFC3 in five transits of planet b and one transit each for planets c and d. The data from planet b did not demonstrate any evidence for an atmosphere \citep{Damiano2022}, and the data collected for planet c was inconclusive, showing modulation in the transmission spectra that could be interpreted as evidence of an atmosphere, but at low significance \citep{barclay2023}. Moreover, there were indications that the signal observed could be a result of contaminating signal from inhomogeneities in the stellar photosphere \citep{barclay2021,barclay2023}. Two upcoming HST/WFC3 transits of planet c may resolve the ambiguity in the results for L~98-59 c. Through Guaranteed Time Observations, JWST will observe L~98-59 c with NIRISS and L~98-59 d with NIRSpec.

% ADDITIONAL DETAILS ON HST AND JWST OBSERVATIONS
% During 2020 and 2021, HST observed a total of 7 transits of this system over 28 orbits (Program ID: 15856). These HST observations were broken up into 5 transits of planet b, and one transit each for planets c and d. Following these observations, an additional HST program (Program ID: 16448) targeting two transits of L~98-59 c was awarded. 
% (GTO). The NIRISS instrument will observe one transit of planet c (Program ID: 1201), and planet d will be observed by NIRSpec over three transits (two transits during Program 1224, and one transit, Program 1201).

%%%%%%%%%%%% TABLE: Initial parameter values  %%%%%%%%%%%%%%%%
\begin{table*} 
\caption{Stellar and planetary system parameters}  \label{tab:initparams}
\centering 
\begin{tabular}{l c l}
\hline\hline
\noalign{\smallskip} 

Parameter & Value adopted for this paper & Source \\
\noalign{\smallskip} 
\hline

{\bf L~98-59} & & \\
Mass (M$_\sun$) & $0.273 \pm 0.030$ & \citet{demangeon2021}\\
Radius (R$_\sun$) &  $0.303 \pm 0.0245$ & \citet{demangeon2021}\\
Temperature (K) & $3415 \pm 135$ & \citet{demangeon2021}\\
XUV saturation fraction & $0.002 \pm 0.001$ & \citet{peacock2020}\\
XUV beta exponent & $0.882 \pm 0.139$ & \citet{peacock2020}\\
XUV saturation time (Myr) & 650 & \citet{peacock2020}\\
Age (Gyr) & $>1$ & \citet{demangeon2021}\\
Min flare energy (erg) & $1.73 \pm 0.39 \times 10^{31}$ & \citet{barclay2023}\\
Max flare energy (erg) & $7.14 \pm 0.88 \times 10^{31}$ & \citet{barclay2023}\\

\hline 
\noalign{\smallskip} 
{\bf L~98-59 b} & & \\
Mass (M$_\earth$) & $0.4 \pm 0.155$ & \citet{demangeon2021}\\
Radius (R$_\earth$) & $0.850 \pm 0.054$ & \citet{demangeon2021}\\
Density (g/cm$^3$) & $3.6$ & \citet{demangeon2021}\\
Flux (S$_\earth$) & $24.7$ & \citet{demangeon2021}\\
Period (days) & $2.2531136$ & \citet{demangeon2021}\\
Semi-major axis (AU) & $0.0218$ & \citet{demangeon2021}\\
Eccentricity & $0.103$ & \citet{demangeon2021}\\
Inclination & 0.0 & \citet{demangeon2021}\\
Argument of periastron (deg) & $192$ & \citet{demangeon2021}\\
Transit time (BJD$_{TDB}$ - 2457000) & $1366.17067$ & \citet{demangeon2021}\\

\hline  
\noalign{\smallskip}                  
{\bf L~98-59 c} & & \\
Mass (M$_\earth$) & $2.22 \pm 0.255$ & \citet{demangeon2021}\\
Radius (R$_\earth$) & $1.385 \pm 0.085$ & \citet{demangeon2021}\\
Density (g/cm$^3$) & $4.57$ & \citet{demangeon2021}\\
Flux (S$_\earth$) & $12.8$ & \citet{demangeon2021}\\
Period (days) & $3.6906777$ & \citet{demangeon2021}\\
Semi-major axis (AU) & $0.0303$ & \citet{demangeon2021}\\
Eccentricity & $0.103$ & \citet{demangeon2021}\\
Inclination & 0.0 & \citet{demangeon2021}\\
Argument of periastron (deg) & $261$ & \citet{demangeon2021}\\
Transit time (BJD$_{TDB}$ - 2457000) & $1367.27375$ & \citet{demangeon2021}\\

\hline
\noalign{\smallskip} 
{\bf L~98-59 d} & & \\
Mass (M$_\earth$) & $1.94 \pm 0.28$ & \citet{demangeon2021}\\
Radius (R$_\earth$) & $1.521 \pm 0.1085$ & \citet{demangeon2021}\\
Density (g/cm$^3$) & $2.95$ & \citet{demangeon2021}\\
Flux (S$_\earth$) & $5.01$ & \citet{demangeon2021}\\
Period (days) & $7.4507245$ & \citet{demangeon2021}\\
Semi-major axis (AU) & $0.0484$ & \citet{demangeon2021}\\
Eccentricity & $0.0740$ & \citet{demangeon2021}\\
Inclination & 0.0 & \citet{demangeon2021}\\
Argument of periastron (deg) & $180$ & \citet{demangeon2021}\\
Transit time (BJD$_{TDB}$ - 2457000) & $1362.73974$ & \citet{demangeon2021}\\

\hline  
\noalign{\smallskip}                  
{\bf All planets} & & \\
XUV absorption efficiency & $0.15-0.3$ & \citet{lugerbarnes2015}\\
XUV absorption radius/planet radius & 1.0 & \citet{lugerbarnes2015}\\

\noalign{\smallskip}
\hline
\hline
\noalign{\smallskip}
\end{tabular}
\end{table*}

 % HOST STAR PROPERTIES  %%%%%%%%%%%%%%%%
\subsection{Host Star Properties} \label{subsec: star prop}

L~98-59 is a small, main-sequence M3 dwarf star with a luminosity about 100 times fainter than our sun \citep[0.012 L$_\sun$;][]{cloutier2019, kostov2019b, demangeon2021}. Due to its slow rotation, \citet{kostov2019b} estimated L~98-59’s age to be $>$1 Gyr. Follow-up calculations by \citet{demangeon2021} agree with this result, stating the age of L~98-59 to be over 800 Myr. This star was also thought to be relatively quiet with little stellar activity, based on lack of H$\alpha$ emission in optical spectra \citep{daria2021, kostov2019b}. However, recent studies have confirmed the detection of flares on L~98-59 \citep{stelzer2022,howard2022,barclay2023}. 

 % PLANETARY SYSTEM PROPERTIES  %%%%%%%%%%%%%%%%
\subsection{Planetary System Properties} \label{subsec: planet prop}

There are four planets in the L~98-59 system, two of which are \added{inferred to be rocky from their bulk densities}\deleted{confirmed to be rocky}. Initially, \citet{cloutier2019} were unable to constrain the mass of L~98-59 b, instead defining an upper limit of 1.01 Earth masses (M$_\earth$). \citet{demangeon2021} were later able to refine the planets' masses to be $0.4^{+0.16}_{-0.15}$, $2.22^{+0.26}_{-0.25}$, and $1.94 \pm 0.28$ M$_\earth$ for planets b, c, and d respectively. As part of their RV analysis, \citet{cloutier2019} also derived densities for L~98-59 c and d, determining planet c to have a rock-dominated composition and planet d to likely have a significant amount of water or gaseous atmosphere due to its lower density. Following up on their results, \citet{demangeon2021} determined that the three inner planets have small iron cores, with densities $3.6^{+1.4}_{-1.5}$ g/cm$^3$ (b), $4.57^{+0.77}_{-0.85}$ g/cm$^3$ (c), $2.95^{+0.79}_{-0.51}$ g/cm$^3$ (d). Their analysis showed that planets b and c have very similar compositions, likely rocky and with a small mass fraction of water and a low gas mass, if any. On the other hand, their model favored a richer gas and water content for planet d. 

\citet{quick2020} reached similar conclusions in their volcanic and cryovolcanic analysis. Using estimates of the planets' total internal heat, which includes both tidal heating and radiogenic heating, they estimated that L~98-59 b and c are likely to exhibit extreme volcanic activity at their surfaces. They also suggested that L~98-59 d may be an ocean world with some volcanic activity, based on the planet's density and effective temperature. 

The L~98-59 system resembles other multi-planet systems around M dwarfs, such as TRAPPIST-1 and Kepler-186 \citep{gillon2017, quintana2014}, in that M dwarfs frequently host compact, multi-planet systems \citep{kostov2019b, muirhead2015}. At 0.0218 AU (b), 0.0303 AU (c), and 0.0484 AU (d), the three inner planets reside interior to both the system's conservative and optimistic HZ, as calculated from \citet{kopparapu2013, kopparapu2014} using the values from Table \ref{tab:initparams}. These HZ limit values are listed in Table \ref{tab:HZ}. The system also follows the ``peas in a pod'' configuration initially described by \citet{weiss2018}, where within a multi-planet system, planets tend to have similar sizes and in systems with more than three planetary bodies, planets tend to be evenly spaced in their orbits.

%%%%%%%%%%%% TABLE: Optimistic & Conservative HZ  %%%%%%%%%%%%%%%%
\begin{table} 
\caption{Optimistic and Conservative Habitable Zones of L~98-59} \label{tab:HZ}
\centering
\begin{tabular}{l c c}
\hline\hline
\noalign{\smallskip}

    HZ Limit & Flux & Distance \\
    \noalign{\smallskip}
    \hline
    \noalign{\smallskip}
    
    \textbf{Conservative} & & \\
    Inner HZ & 0.9298 S$_\earth$ & 0.109 AU \\ 
    Outer HZ & 0.244 S$_\earth$ & 0.212 AU \\
    \noalign{\smallskip}
    
    \textbf{Optimistic} & & \\
    Inner HZ & 1.492 S$_\earth$ & 0.086 AU \\
    Outer HZ & 0.22 S$_\earth$ & 0.224 AU \\

\noalign{\smallskip}
\hline
\end{tabular}
\end{table}

 %%%%%%%%%%%%%%%%
 % NUMERICAL METHODS
 %%%%%%%%%%%%%%%%
\section{Numerical Methods and Models} \label{sec:model}

% REMOVED -- SYSTEM STABILITY MODEL -- REMOVED

%\subsection{System Stability Model}\label{sec:stability}

\deleted{The stability of a planetary system can give important constraints on the measurements of the bodies' parameters; in the case of L~98-59, we assume the system to be stable with the parameters we adopt in this paper. We verified system stability with the software package \texttt{mercury6} \citep{chambers1999}. \texttt{Mercury6} is an N-body integrator capable of calculating the orbital evolution of a set of bodies around a larger central body. We derive the dynamics equations from \citet{murray1999solar}:}

% x component of velocity vector for dynamics equation
%\begin{equation}
\deleted{$\dot{x} = -\frac{n a}{\sqrt{1-e^2}}sin(f)$}
%\end{equation}

% y component of velocity vector for dynamics equation
%\begin{equation}
\deleted{$\dot{y} = +\frac{n a}{\sqrt{1-e^2}}(e+cos(f))$}
%\end{equation}

\deleted{\noindent where $n$ is the mean motion, $a$ is the semi-major axis, $e$ is the eccentricity, and $f$ is the true anomaly. With these equations, we determine the position and velocity of each planet, which the integrator takes as its initial parameters, using the astronomical coordinates of the L~98-59 planets in Table \ref{tab:initparams}. We perform an evolution over $10^5$ years, and find that the L~98-59 system remains stable throughout.}

 % SYSTEM EVOLUTION MODEL  %%%%%%%%%%%%%%%%
\subsection{System Evolution Model} \label{sec:evol_model}

To study the atmospheric development of the L~98-59 planets, we simulate the evolution of the system over several billion years assuming an initial water content between 1 and 100 terrestrial oceans (TO) and accounting for the host star's radiation and flares. We use the modeling software package \texttt{VPLanet} \citep[Virtual Planet Simulator;][]{barnesvpl} to model the atmospheric escape of the three innermost, terrestrial planets of the L~98-59 system. \texttt{VPLanet} is an open-source software for simulating the evolution of a planetary system, focusing specifically on habitability. By coupling various models, it allows users to incorporate specific modules and features into their planetary system, such as tidal evolution or a circumbinary system. We will discuss each module we use in our model in Sections \ref{subsec:stellar} (\texttt{STELLAR}), \ref{subsec:flare} (\texttt{FLARE}), and \ref{subsec:atmesc} (\texttt{AtmEsc}). We evolve the system from before the star's main sequence (MS) phase (5 Myr) until the upper limit for the star’s age at the present day (13 Gyrs), in order to visualize the full timescale of this system’s evolution\added{, which has previously been shown to be dynamically stable \citep{kostov2019b, demangeon2021}.} 

We make three important assumptions in the initial setup of our model. First, we assume that the planets are in their currently-observed orbits at the start of each simulation. We thus begin the simulations with the planets in their present orbits around L~98-59, and assume that any planet migration in the early stages of the system's evolution has already taken place. This first assumption ties into our second one, in which we begin the simulations before the star's main sequence phase, at a stellar age of 5 Myr. Pre-MS M dwarfs emit higher amounts of flux in the XUV as compared to MS stars \citep{ramirez2014, lugerbarnes2015, tianida2015, doamaral2022}. Excess XUV flux can cause more water loss and oxygen accumulation early in a planet's evolution, so it is important to include that early evolution time in our simulations; by not doing so, we might severely underestimate the desiccation of the L~98-59 planets. Third, we assume that all three planets start with the same water mass at the beginning of each simulation. This allows us to directly compare the effects of L~98-59's stellar activity and evolution on each planet in the system, and in turn, how water content changes on each of them. 

We investigate the evolution and outcome of three different atmospheric escape scenarios by varying the initial water content of the L~98-59 planets. \citet{doamaral2022} found that the amount of escaping water is independent from the starting quantity of water on a planet, and that the star mass is inversely correlated with the amount of water lost. In other words, planets tend to lose the same amount of water regardless of their initial budget, and smaller stars will contribute most to this water loss due to their longer pre-main-sequence (PMS) phase. Given we do not know the underlying initial water budget for the L~98-59 planets, we vary the starting water budgets, in order to gain a broader understanding of the water retention on each planet over time.

We vary the water budgets, running sets of simulations with initial water masses of 1, 10, and 100 TO following the procedures of \citet{morbidelli2000, raymond2006, chassefiere2012, lugerbarnes2015, doamaral2022} in order to account for the possibility of the planets forming in situ, or migrating in from further out in the disk where more water may have been available at planet formation \citep{mordasini2012, tianida2015, unterborn2018, daria2021}. Our parameter variation method is further discussed in Section \ref{subsec:paramVar}.
        
We describe the specific modules used in our model in the following subsections. Using the \texttt{VPLanet} modules \texttt{STELLAR} and \texttt{FLARE}, we account for both XUV and flare activity from the L~98-59 star in our model. We also use the \texttt{AtmEsc} module to simulate the planets' atmospheric and oceanic evolution over time.

 % STELLAR MODULE  %
\subsubsection{\texttt{STELLAR} module} \label{subsec:stellar}

As a star evolves and ages, the changes in its stellar activity affect the formation, evolution, and potential habitability of its surrounding planets. Early M stars, such as L~98-59, emit near-constant elevated levels of XUV flux for the first few hundred million years before decreasing with time ($\sim t^{-1}$) by two orders of magnitude towards field age \citep{peacock2020}. This decrease in emission is linked to the spin down of the star reducing the dynamo production of its magnetic field \citep{west2015}.

The \texttt{STELLAR} module allows users to include the star's changing parameters into the system evolution model, including its XUV parameters and stellar evolution model \citep{baraffe2015}. Following the results from \cite{peacock2020}, we adopt an XUV saturation timescale of 650 Myr for L~98-59. To determine the XUV saturation fraction and $\beta$ exponent for L~98-59, we use evolutionary models of 0.35 M$_\sun$ stars from \cite{peacock2020}\footnote{\url{https://stdatu.stsci.edu/hlsp/hazmat}} in combination with \textit{The R\"{o}ntgensatellit} (\textit{ROSAT}) X-ray measurements of proxy stars from \cite{shkolnik2014} to create full panchromatic spectra representative of L~98-59 at ages between 10 Myr and 5 Gyr\footnote{5 Gyr is the oldest spectrum available from \cite{peacock2020}. At X-ray wavelengths (1-100 \AA), we adopt a single value flux consistent with the median \textit{ROSAT} measurement for each given age bin. Since the XUV flux declines in a consistent manner beyond 650 Myr, calculating the XUV saturation fraction and $\beta$ exponent with models between 650 Myr and 5 Gyr (rather than 13 Gyr) yields a sufficiently long baseline to use.}. We use the \cite{peacock2020} models for wavelengths $>$100 \AA\ and the \textit{ROSAT} measurements for wavelengths $<$100 \AA. We compute the XUV (1 -- 1000 \AA) and bolometric luminosities from these spectra and fit a power law to the decreasing L$_{XUV}$/L$_{bol}$ to determine the $\beta$ exponent (the coefficient in this power-law fit). We do this with the median models for each age and the inner quartile models to provide an uncertainty: $\beta$ = $0.882 \pm 0.139$. The XUV saturation fraction is L$_{XUV}$/L$_{bol}$ at the saturation time of 650 Myr, L$_{XUV}$/L$_{bol}$ = $0.002 \pm 0.001$.

 % FLARE MODULE  %
\subsubsection{\texttt{FLARE} module} \label{subsec:flare}

The presence of flares in a planetary system can have a strong impact on the habitability of a planet, particularly on the retention and detection of its atmosphere. Flares can alter the chemical composition of a planet's atmosphere, and repeated exposure to flares can cause a planet to quickly lose any atmosphere it might have accumulated \citep{Howard2021flareatm, Tilley2019repflare, louca2023}. Until now, little work has been done to study the impacts of flare exposure on a planet's ocean retention, even though the most commonly observed stars for small, rocky exoplanet characterization -- main sequence M stars, or M dwarfs -- are known to emit such radiation \citep{billings2011, shields2016, fujii2018, doamaral2022}. Our evolution model thus includes the effects of flares on the atmospheric escape and ocean retention of the L~98-59 planets.
    
The \texttt{FLARE} module in \texttt{VPLanet} enables users to specify basic information on the star's flare distribution. We identify 8 flares in Sector 11 of TESS observations using a modified version of \texttt{bayesflare} \citep{pitkin14} as described in \citet{gilbert2022}. Using the identified flare parameters (peak time, amplitude, FWHM) as identified in the flare detection, we model these flares using \texttt{xoflares} \citep{xoflares}. We integrate under these models to determine the equivalent duration of each flare, then scale these to absolute energies using the convolution of an M3 spectral model of L~98-59 and the TESS bandpass. From here, we determine the mean energy of the star and then use the model of the individual flares from \citet{barclay2023} to determine the minimum and maximum flare energies, which are $1.73^{+0.38}_{-0.41} \times 10^{31}$~erg and $7.14^{+0.88}_{-0.88} \times 10^{31}$~erg, respectively. These uncertainties are averaged such that we use a minimum flare energy of $1.73 \pm 0.39 \times 10^{31}$~erg and a maximum flare energy of $7.14 \pm 0.88 \times 10^{31}$~erg. We use the ``TESSUV'' option for the ``sFlareBandPass'' parameter, which follows \citet{gunther2020} to calculate which fraction of the bolometric energy for observed flares fall into the U-band (E$_U\approx$ 7.6\%E$_{\rm bol}$).

 % ATMESC MODULE  %
\subsubsection{\texttt{AtmEsc} module} \label{subsec:atmesc}

Atmospheric escape is the process undergone by a planet when gases from its atmosphere are lost to space. There are two broad categories of atmospheric escape defined by the gas loss mechanism: thermal escape, and non-thermal escape. Thermal escape includes Jeans escape, in which the temperature of a gas accelerates its particles to above the escape velocity, and hydrodynamic escape, in which escaping gas molecules drag along other molecules, creating a fluid-like escape behavior. Non-thermal escape includes processes such as photochemical loss, ion loss, and ionospheric outflow (among others), and generally involves more complex interactions in ions and plasma (see \citet{gronoff2020} for more detail on atmospheric escape processes).

Thermal escape is considered the dominant mechanism for highly irradiated planetary atmospheres, so the \texttt{VPLanet} model includes energy-limited and diffusion-limited escape for H/He and water vapor atmospheres. Energy-limited escape occurs as a hydrodynamic wind ``blows away'' hydrogen in the atmosphere, and is driven by a fixed fraction of the incoming XUV energy. However, if not all of the oxygen escapes and some remains in the atmosphere, the hydrogen will have to escape by diffusing through a background of oxygen in the atmosphere, thus causing diffusion-limited escape \citep{lugerbarnes2015}. 
    
The \texttt{AtmEsc} module in \texttt{VPLanet} describes the evolution of a planet's atmosphere and ocean retention. We specify each planet's starting water budget as 1, 10, or 100 TO, and vary this quantity in a way that is further discussed in Section \ref{subsec:paramVar}. We follow the water loss model described in \citet{lugerbarnes2015} to simulate the evolution of water on each planet and to set some of our parameters: we set the XUV absorption radius to be equal to the planet radius, since XUV radiation is absorbed in the uppermost layers of a planet's atmosphere, and we set the XUV absorption efficiency parameter to vary between 0.15 - 0.3. We select the model from \citet{bolmont2017} in the \texttt{VPLanet} code to determine the XUV absorption efficiency parameter for water vapor. In these simulations, we assume minimal O$_2$ absorption at the surface \citep{lincowski2018}, so we do not include O$_2$ loss processes other than the ones built into the \texttt{VPLanet} integrator. We also assume no significant H/He envelopes. We keep the Jeans Time, in which the ballistic escape of individual atoms drives atmospheric mass loss in the low temperature limit \citep{luger2015}, to the default value of 1 Gyr.

% PARAMETER VARIATION  %
\subsubsection{Parameter Variation} \label{subsec:paramVar}

In order to consider the full range of scenarios for the system’s evolution, we vary the following parameters within their uncertainties: star mass (M$_s$), planet mass for all three planets (M$_p$), planet radius for all three planets (R$_p$), XUV saturation fraction, XUV beta exponent, minimum flare energy, and maximum flare energy. Using Monte Carlo sampling, we draw 1000 values from a Gaussian distribution for each of these parameters. We also include each planet's XUV absorption efficiency in our sampling as a flat distribution from 0.15 to 0.3, which is considered the range appropriate for planets with hydrogen-rich atmospheres around M dwarfs \citep{lugerbarnes2015}. We then use these varying parameters to generate 1000 unique simulations, spanning the parameter space of the L~98-59 system. We will refer to these 1000 simulations as a \textit{set} of simulations. We fix all other parameters in the model, excluding the semi-major axis which varies indirectly due to \added{Newton's version of }Kepler's third law.

We perform a set of simulations for three initial water masses: 1, 10, and 100 TO on each planet. For each initial water budget, we simulate the instellation and each planet's climatic response using input parameters as described above to build a distribution of water loss and atmospheric escape scenarios. In our model, we consider ``desiccated'' to mean that a planet's water content has dropped \added{to zero. We note that this may affect the interpretation of our results, as even simulations that retain insignificantly small amounts of water are still considered non-desiccated.}\deleted{below a significant level, which we set to 0.1 TO; any water below that quantity is too low for us to consider it significant in the scheme of JWST observations.} Thus, we model 1000 total scenarios per starting water budget, with which we determine the range of most likely scenarios within 3$\sigma$ for water retention and oxygen accumulation on the L~98-59 planets. That is, given the distribution of our simulation results, we find the water evolution scenario at +3$\sigma$ to represent the ``most'' or ``longest'' water retention and the water evolution scenario at -3$\sigma$ to represent the ``least'' or ``shortest'' water retention, giving us a range of most likely water and oxygen evolution behaviors for each planet.

%--------------------------------------------------------------
%%%%%%%%%%%%%%%%%%%%%%%%%% MAIN PLOTS %%%%%%%%%%%%%%%%%%%%%%%%%%
%--------------------------------------------------------------

% Water and oxygen evolution plots for each planet:

% 1 TO
\begin{figure*}[ht]
\centering
    \includegraphics[width=1.0\textwidth]{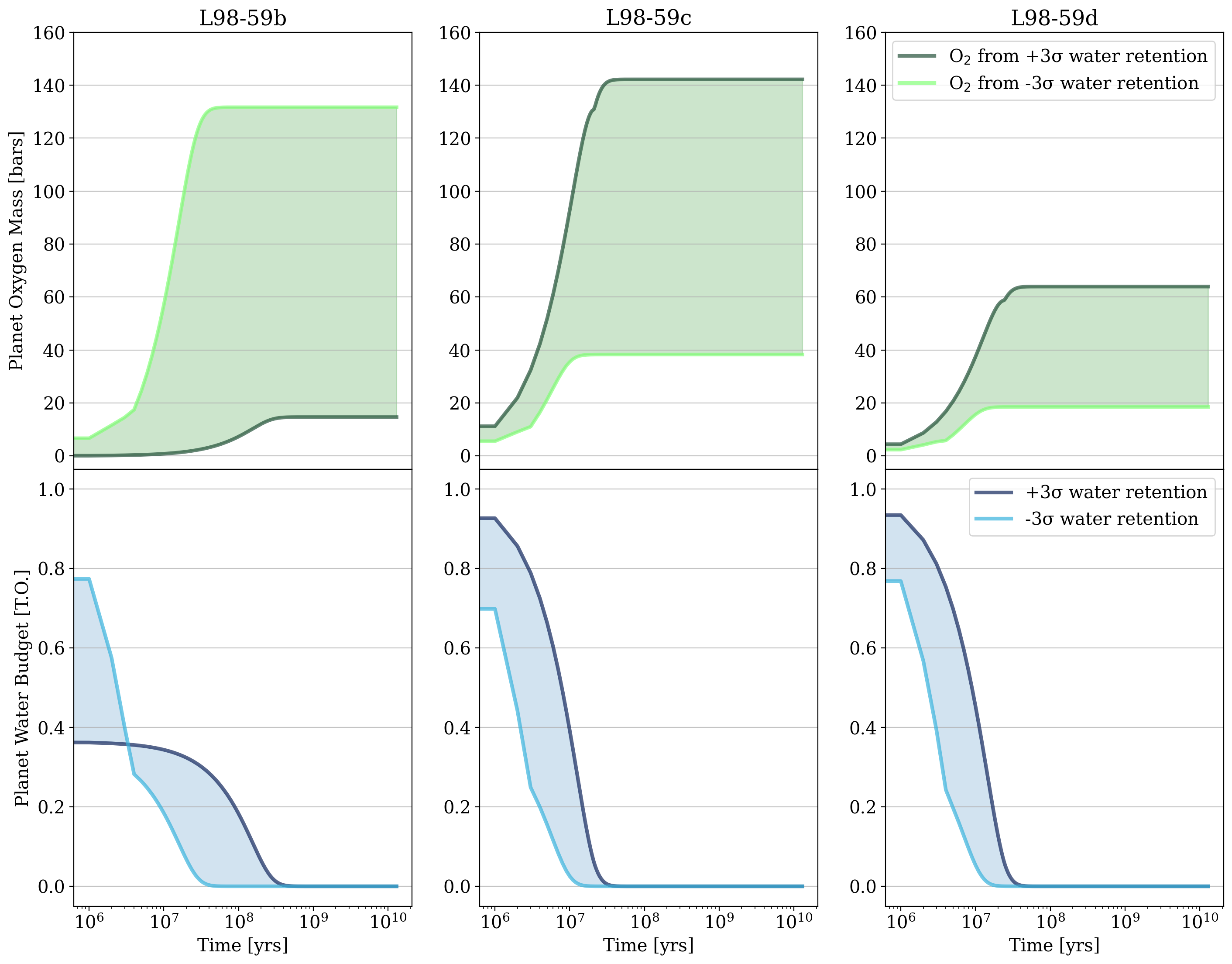}
    \caption{\textbf{1 TO} These plots show the range of outcomes for oxygen accumulation (top row, green) and water retention (bottom row, blue) for the set of simulations with 1 TO initial water content on planets b (left column), c (middle column), and d (right column). The dark and light blue lines represent the simulations corresponding to the +3$\sigma$ and -3$\sigma$ (respectively) of the water desiccation times distribution for each planet at 1 TO. These distributions can be found in Figure \ref{fig:combined_hists}, top row. The dark and light green lines correspond to the same simulations as the ones chosen in the water behavior plots. Given 1 TO, water isn't likely to survive past 1 Gyr for any planet, which makes water observations more unlikely. However, the quantities of oxygen produced are still significant, up to \added{140}\deleted{100} bars. Notice that for planet b, simulations that retain water longer are the ones that accumulate the least oxygen, contrary to planets c and d. These b simulations also show a quick initial drop in water content, and then a slower loss over time than the simulations that retain water for shorter times. This is due to a high XUV flux for b, which increases radiative cooling and decreases escape efficiency. Further detail is provided in Section \ref{sec:results}.} 
    \label{fig:best_worst1}
\end{figure*}

% 10 TO
\begin{figure*}[ht]
\centering
    \includegraphics[width=1.0\textwidth]{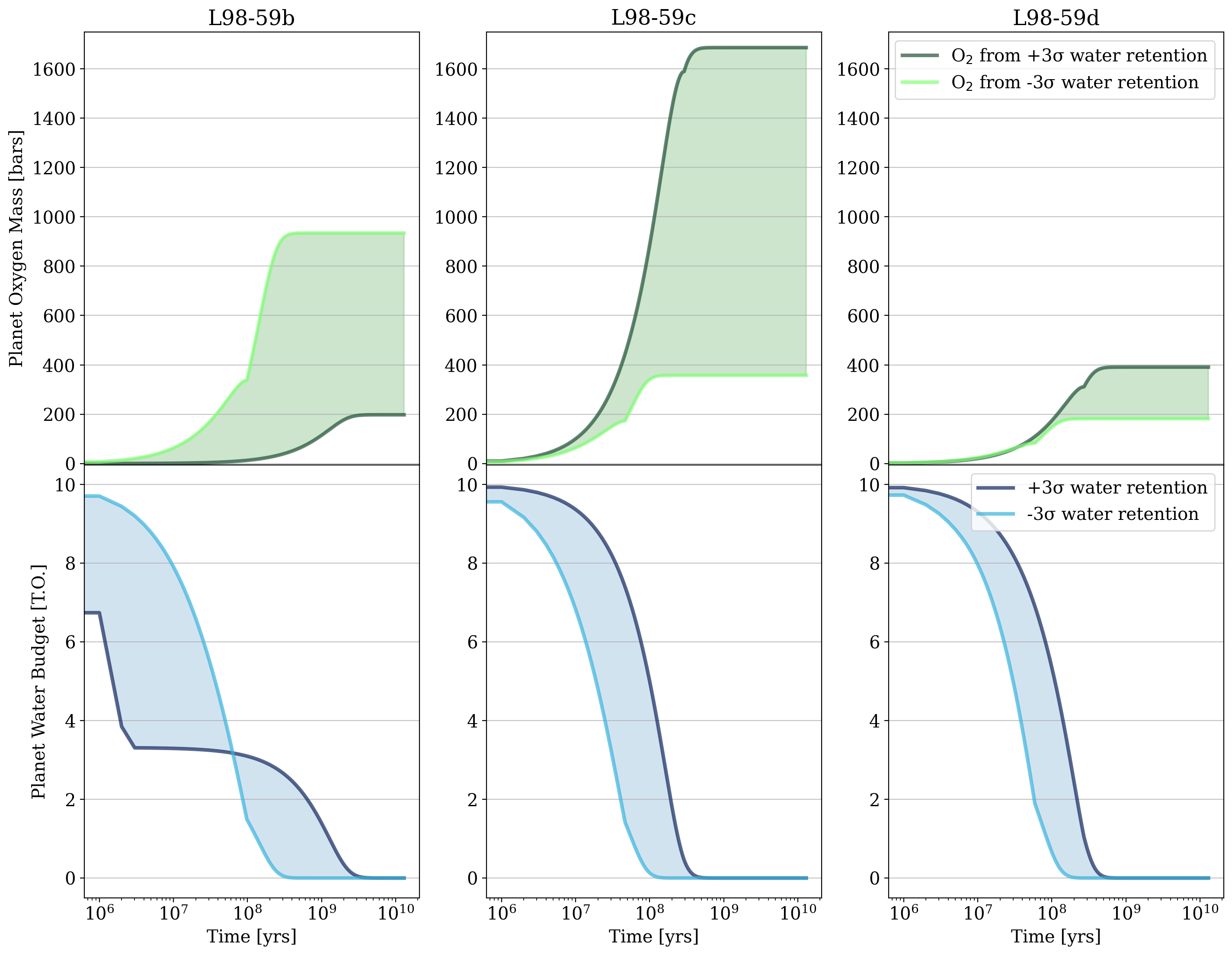}
    \caption{\textbf{10 TO} These plots show the range of outcomes for oxygen accumulation (top row, green) and water retention (bottom row, blue) for the set of simulations with 10 TO initial water content on each of the planets. The distributions for the water behavior can be found in Figure \ref{fig:combined_hists}, middle row. Given 10 TO, all three planets can accumulate hundreds of bars of oxygen, and planet b is likely to retain water for longer than planets c and d (\added{likely }past 1 Gyr) making water detections more likely for this planet. However, depending on the system's age, all three planets may still have some water left. Details on planet b's behavior, similar to Figure \ref{fig:best_worst1}, are given in Section \ref{sec:results}.}
    \label{fig:best_worst10}
\end{figure*}

% 100 TO
\begin{figure*}[ht]
\centering
    \includegraphics[width=1.0\textwidth]{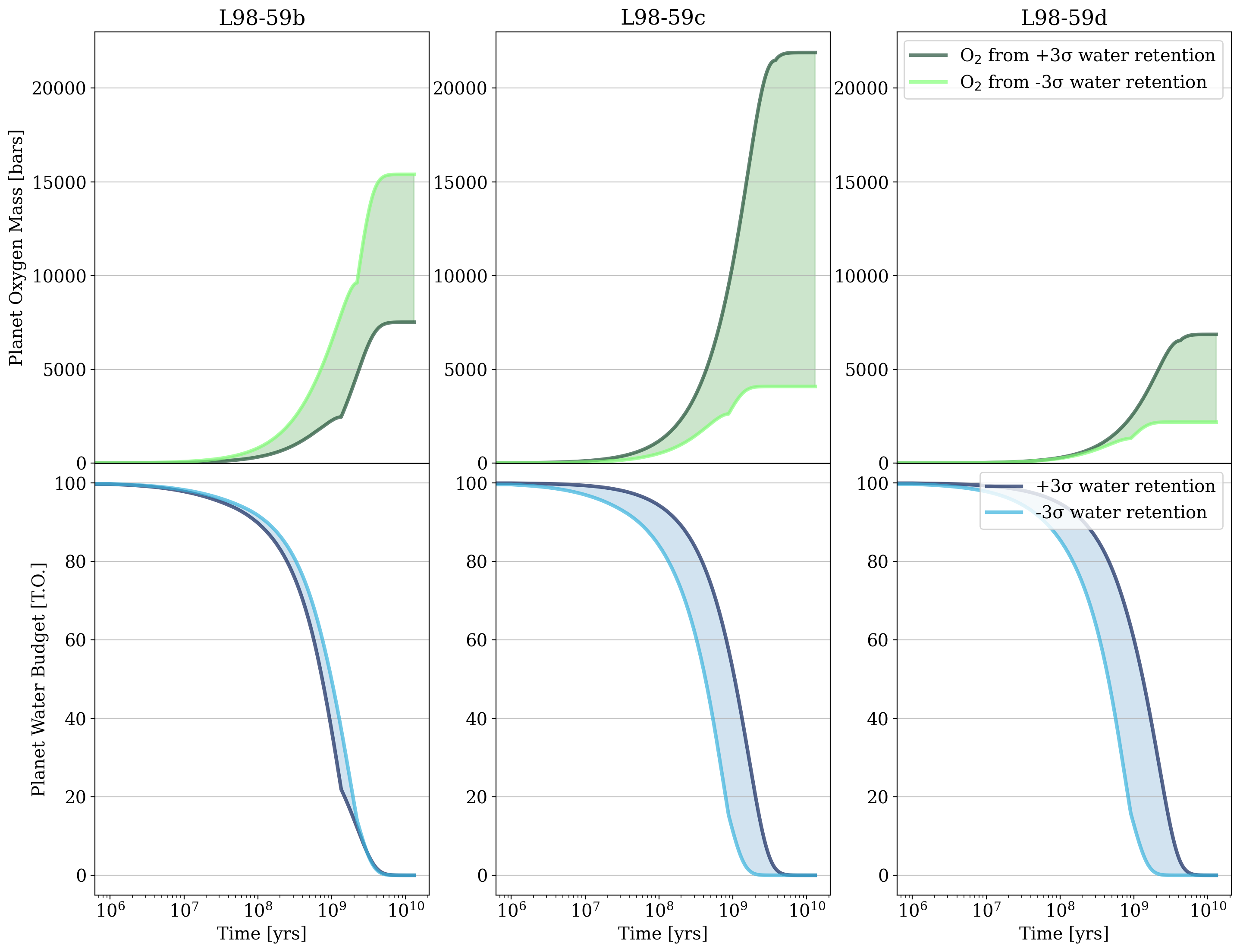}
    \caption{\textbf{100 TO} These plots show the range of outcomes for oxygen accumulation (top row, green) and water retention (bottom row, blue) for the set of simulations with 100 TO initial water content on each of the planets. The distributions for the water behavior can be found in Figure \ref{fig:combined_hists}, bottom row. Given 100 TO, all three planets are likely to retain water past 1 Gyr, making observations of water relatively likely. In addition, thousands of bars of oxygen may be produced for each planet in this scenario. Details on planet b's behavior, similar to Figure \ref{fig:best_worst1}, are given in Section \ref{sec:results}.}

    \label{fig:best_worst100}
\end{figure*}

 %%%%%%%%%%%%%%%%
 % RESULTS
 %%%%%%%%%%%%%%%%
\section{Results} \label{sec:results}

% MAIN RESULTS SUMMARY
The main results from our simulations, showing the water retention and oxygen accumulation over time for the L~98-59 planets at different starting water budgets, are summarized in Figures \ref{fig:best_worst1}, \ref{fig:best_worst10}, and \ref{fig:best_worst100}. We find that all three planets are experiencing strong atmospheric escape due to their proximity and exposure to their star's XUV and flare activity. All three planets are able to accumulate significant quantities of oxygen in their atmospheres over time. While all three planets display significant water loss, planet b shows the potential for water retention when given a medium to high initial water content, but planets c and d are unlikely to retain significant amounts of water unless given a high initial water content.
    
\begin{figure*}[ht]
\centering
    \includegraphics[width=1.0\textwidth]{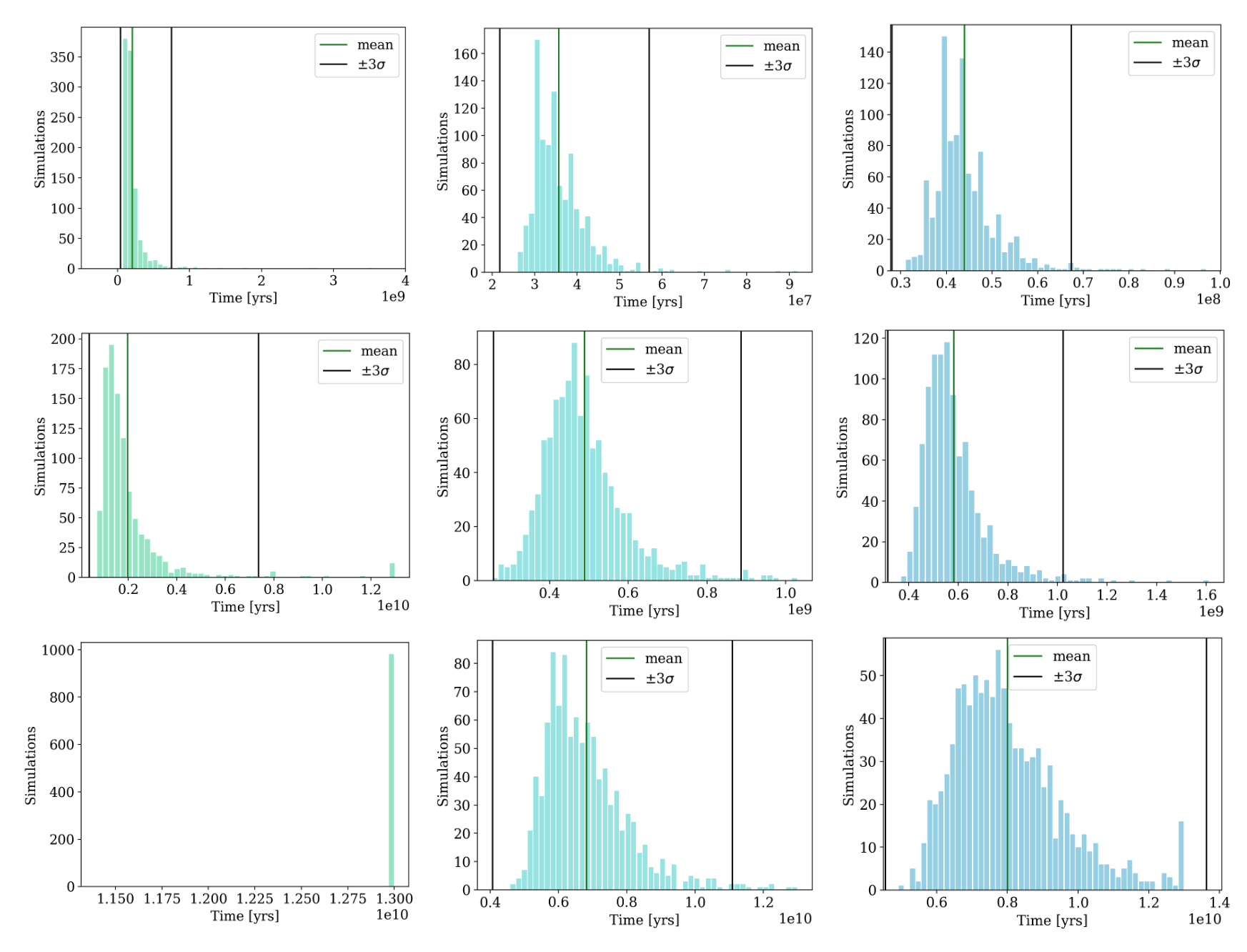}
    \caption{These plots show the distribution of the \added{water }desiccation times for the set of simulations run at 1 TO (top row), 10 TO (middle row), and 100 TO (bottom row) for planets b (left column, green), c (middle column, light blue), and d (right column, dark blue)\added{, spread out over 50 bins}. The vertical green line and the vertical black lines in each plot show the log mean and $\pm3\sigma$ of the results distribution, respectively. In all three initial water cases, planet b on average retains water the longest. In the bottom left histogram, the fullest bin on the far right of the plot shows all cases of water survival for planet b at 100 TO. Note that \deleted{this is not necessarily the most common case: }the cases of water survival only \added{accumulate in the 1.3 Gyr bin}\deleted{show themselves as times beyond 1.3 Gyr}, as the simulations all stopped at time 1.3 Gyr. \added{Although it seems like all the simulations retained water in planet b's 100 TO case, a few simulations being obscured by the magnitude of the surviving cases were fully desiccated. }A distribution of only the desiccated \deleted{times} \added{cases and a distribution of only the water surviving cases} for planet b can be found in Figure \ref{fig:100TO_B_desicc}.}
    \label{fig:combined_hists}
\end{figure*}

\begin{figure*}[ht]
\centering
    \includegraphics[width=0.8\textwidth]{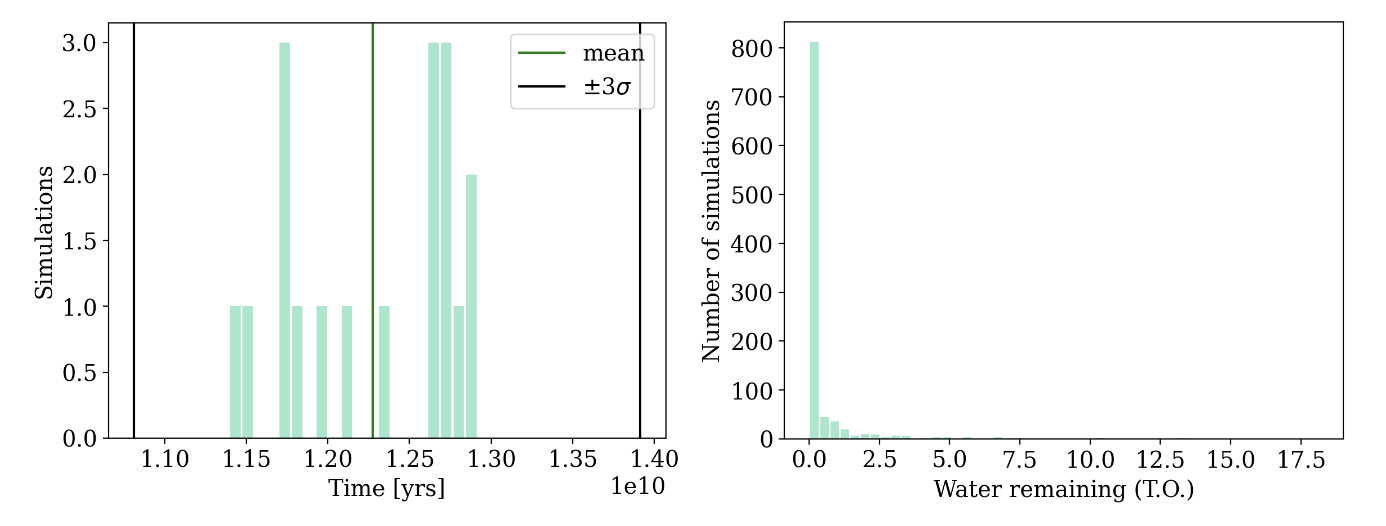}
    \caption{This plot shows the distribution of \deleted{\textit{only}} the desiccated cases \added{(left) and the surviving water cases (right)} for the set of simulations run at 100 TO for planet b, as seen on the bottom left plot of Figure \ref{fig:combined_hists}. \added{The desiccated cases plot shows when simulations ran out of water before 13 Gyr, spread over 20 bins, and the surviving water cases shows how much water remained in simulations that survived past 13 Gyr. We use different bin numbers for both plots because very few cases fully ran out of water before the end of our simulation time. }The vertical green line and the vertical black lines \added{on the left plot} show the log mean and $\pm3\sigma$ of the results distribution, respectively. \added{Note that although planet b desiccated in very few simulations given 100 TO, most cases that made it past 13 Gyr still had very little water remaining. However, it}\deleted{It} is interesting to consider that, depending on the L~98-59 system's age, planet b may still have some water present.}
    \label{fig:100TO_B_desicc}
\end{figure*}

% planet b might retain some water given enough starting water:
As seen in the histograms from Figure \ref{fig:combined_hists}, when given 100 TO of starting water, \added{most}\deleted{more than a quarter} of the simulations predict that planet b will not desiccate and will instead retain some fraction of water, as opposed to planets c and d which still tend to desiccate in this situation. At higher XUV fluxes, such as ones that planet b experiences, the escape efficiency decreases due to increased radiative cooling. This means that less hydrogen is escaping, but that which is escaping leaves the atmosphere at a high velocity and drags along large amounts of oxygen with it \citep{tian2008, bolmont2017}. \added{Note that although planet b retains water in most cases given 100 TO, the surviving simulations still likely have very little water left over (see Figure \ref{fig:100TO_B_desicc}.}

% planets c/d unlikely to retain much water, unless given lots of starting water
Planets c and d are both less likely than planet b to retain significant amounts of water over long periods of time, unless they begin evolving with high quantities of water. Indeed, most of our \texttt{VPLanet} simulations predict that in cases of 1 TO and 10 TO of initial water, planets c and d completely desiccate before 1 Gyr, as seen in Figures \ref{fig:best_worst1}, \ref{fig:best_worst10}, \added{and \ref{fig:combined_hists},} unless they start their evolution with a significant amount of water. Given 100 TO, planet c is able to retain water for close to \added{11}\deleted{6} Gyr (Figure \ref{fig:combined_hists}, bottom middle), and planet d shows a few cases of retaining water for more than 13 Gyr (Figure \ref{fig:combined_hists}, bottom right). \deleted{However, those cases are rare and significantly outside of our $\pm 3\sigma$ boundaries for water desiccation times, so we consider them to be outliers which would not accurately reflect the current observable state of planet d.}

% all planets retain a lot of oxygen!
The atmospheric escape resulting from the star's intense XUV radiation causes the planets to accumulate significant quantities of oxygen, in both low and high initial water content cases. Given 1 TO, the planets can accumulate from around \added{15}\deleted{$20$} bars up to more than \added{140}\deleted{100} bars of O$_2$. In addition, given 100 TO the planets can accumulate \deleted{hundreds to }thousands of bars of O$_2$. Unless the planets began their orbits completely desiccated and stripped of any volatiles, there is a strong chance of observing secondary atmospheres dominated by oxygen on all three of them. We note that these simulations do not take into account potential volcanic activity on the planets, which as discussed in Section \ref{subsec: planet prop}, is a likely scenario for planets b and c. This implies that the planets may also accumulate other volatiles produced by volcanic outgassing, besides O$_2$. Our simulations also do not account for potential sinks, like magma oceans -- this is further discussed in Section \ref{subsec:limits}.

% behavior of b and c/d is switched for oxygen and water!
Examining the water evolution displayed in Figures \ref{fig:best_worst1}, \ref{fig:best_worst10}, and \ref{fig:best_worst100}, we find that the water loss and oxygen accumulation over time consistently do not behave the same way for planet b as they do for planets c and d. For planet b, the simulations that seem to retain water for shorter periods of time have a rather fast and steady water loss rate right from the beginning, which only slows down when the planet reaches insignificant levels of water. However, the simulations that seem to retain water for longer begin their evolution with an instantaneous drop in water quantity, then continue with a much slower water loss rate. These simulations also accumulate much less oxygen than simulations with shorter water retention times. By contrast, planets c and d display a more constant water loss rate in all simulations throughout the sets, and their simulations that retain water the longest also accumulate the most oxygen. 

This opposing behavior in water loss and oxygen retention is related to how the escape efficiency depends on the XUV flux, for the same reason why planet b is more likely than the other planets to retain some water, given a high initial water content. Planet b experiences a higher XUV flux than the other planets, such that it has a lower escape efficiency and could then retain water for longer. But with a high XUV flux, the hydrolyzed oxygen that does accumulate in the atmosphere will be dragged away by quickly escaping hydrogen. Planets c and d, with a higher escape efficiency, will experience a steeper and more constant water loss. Exposure to a less intense XUV flux also means that they can accumulate more oxygen, since it does not get carried away by hydrogen as dramatically as for planet b.

% Combined histograms of all planets at all TO.

%%%% Desiccated cases for planet b at 100 TO

 %%%%%%%%%%%%%%%%
 % DISCUSSION
 %%%%%%%%%%%%%%%%
\section{Discussion} \label{sec:discuss}

% Venus nature of the planets (runaway greenhouse, lose water, gain volailes)
\subsection{L~98-59 planets are in a runaway greenhouse}

The L~98-59 planets have mean semi-major axes values of 0.0218 AU (b), 0.0303 AU (c), and 0.0484 AU (d), and thus have instellations ranging from about 4--25 times the instellation that the Earth receives from the Sun (Figure~\ref{fig:planet_rad}). These planets therefore fall firmly in the Venus zone \citep{kane2014, Ostberg2023}, the region around a star within which a planet like Earth with water on its surface would likely have been forced into a runaway greenhouse. Most Venus analogs have been found around faint stars for which it is difficult to obtain follow-up measurements. However, the L~98-59 planets' location along with the brightness of their host star create a convenient combination which would allow us to learn more about the evolution and atmospheric development of Venus-like planets \citep{daria2021}. L~98-59 c and d, if they have atmospheres, are prime JWST targets to observe and study potential Venus analogs and further study the origin and backstory of our sister planet \added{\citep{kane2019}}.

As seen in Figures \ref{fig:best_worst1}, \ref{fig:best_worst10}, and \ref{fig:best_worst100}, the L~98-59 planets are likely experiencing a runaway greenhouse, which makes them an excellent target for JWST observations of evolving atmospheres and unstable climates. In a runaway greenhouse state, the planets' surface temperatures exceed the critical point of water (647 K) and their oceans start evaporating \citep{kastingh1988, kasting1993, abe1993}. Our simulation results show that all three planets are constantly losing water and accumulating high quantities of oxygen from water photolysis, because they are so small and so close to their star. \citet{schindlerkasting2000} modeled synthetic spectra of hypothetical Earth-like atmospheres, which showed that ``Venus-like'' planets could conceivably accumulate large amounts of oxygen in their atmospheres. \citet{lugerbarnes2015} confirmed this behavior in their planetary evolution and water loss model, demonstrating that planets around the habitable zone of M dwarf stars may develop atmospheres with hundreds to thousands of bars of O$_2$. Thus, if the three planets had started their evolution with any non-negligible water budget, it is likely that they would display signs of water loss and volatile accumulation in their atmospheres. Any detection of volatiles on the L~98-59 planets would show an impermanent state in the lifetime of the system, but would also offer real-time snapshots of a runaway greenhouse state and its effects on the climate stability and evolving atmospheric composition of a planet.

% Comparing to recent flat spectrum
\subsection{Potential future volatile detections} \label{subsec:discussion_volatiles}

Depending on the system's age, there is a real possibility of observing potential water signatures on L~98-59 b, as it seems prone to longer water retention than L~98-59 c and d. Additionally, our simulations predict a significant accumulation of oxygen in planet b's atmosphere, especially in the case of 100 TO. Such quantities of oxygen and water vapor, resulting from water loss, may create a haze in the planet's atmosphere, thus obscuring potential water signal detections. Indeed, a recent transmission spectrum of L~98-59 b, found to be relatively flat, predicts that planet b either has no atmosphere or an atmosphere with high-altitude clouds or haze \citep{Damiano2022}. The authors' data rule out a cloud-free and H$_2$-dominated atmosphere, which is consistent with our initial assumptions about the planet. However, this transmission spectrum does not favor a water vapor-dominated atmosphere either, although the possibility remains if clouds are included in the model. 

JWST observations of planet b in parallel with planets c and d would provide essential information on the evolution of different planets around the same M dwarf. Not only would it allow us to compare the development and loss of atmospheres on three small planets in the same system, but the detection of H$_2$O on L~98-59 b would allow us to place constraints on evolutionary scenarios, by looking at how oceans and atmospheres behave when subjected to such intense irradiation. For example, we may determine whether these planets formed with a large initial amount of water, giving us a way to probe into the mechanism of volatile accumulation during the initial formation stage of planets.

\added{Thanks to the sensitivity of its instruments, JWST would be capable of detecting the presence of oxygen on the L~98-59 planets. Planets b and d of the system are both listed in the best-in-class planet candidates list compiled by \citet{hord2023} for having the potential to yield high quality spectra. This list accounts for the sensitivity of  JWST's instruments and shows that these L 98-59 planets have a relatively high Transmission Spectroscopy Metric \citep[TSM;][]{kempton2018}. Additionally, \citet{daria2021} perform simulations using the Planetary Spectra Generator \citep[PSG;][]{villanueva2018} to determine the detectability of different atmospheric features on the L~98-59 planets by JWST. Specifically they predict that NIRISS SOSS, and even MIRI LRS, would be capable of detecting the apparent features of a desiccated O$_2$ atmosphere (see Figures 8 and 9 from \citet{daria2021} for further detail). Although ozone (O$_3$) can result from the photochemical processing of O$_2$, the L~98-59 planets should have hot enough atmospheres that O$_3$ production would be significantly reduced, making it unlikely that these features would interfere with potential O$_2$ detections. However, we note that our results predicting high quantities of oxygen on the L~98-59 planets represent an \textit{upper limit} on the accumulation of oxygen. Because the potential presence of oxygen sinks on the planets would decrease the available atmospheric oxygen, we should interpret these results as describing how much oxygen the XUV radiation of a star can produce on volatile-rich planets, rather than how much oxygen will be available to detect.}

% False biosignatures, flares
\subsection{Assumptions regarding flares and water delivery}

As an M dwarf, L~98-59 is more prone to stellar variability and flares, which presents a challenge for the planets' ability to retain any significant atmosphere or ocean. XUV radiation emitted by M dwarfs heats the exosphere of a planet's atmosphere, causing the exobase to expand, and therefore facilitating escape \citep{murrayclay2009, france2020}. Thus, continuous exposure to repeated flaring can significantly deplete an atmosphere \citep{Tilley2019repflare, france2020}. It is likely that any water on the L~98-59 planets would be significantly affected, if not desiccated, by the star's flares and intense XUV radiation, as water-rich atmospheres may be severely depleted during a star's PMS phase \citep{tianida2015, lugerbarnes2015, doamaral2022}.

In addition, flares can generate up to hundreds of bars of additional O$_2$ due to water photolysis \citep{doamaral2022}, and planets in an extended runaway greenhouse in the HZ of M dwarfs can also accumulate hundreds to thousands of bars of O$_2$ \citep{lugerbarnes2015}. So while these atmospheric escape processes might create detectable amounts of oxygen on the L~98-59 planets, we note that this would be a false biosignature and an unreliable measure of the habitability of these planets \citep{wordsworth2014, tian2014, domagalgoldman2014, lugerbarnes2015}. Although M dwarfs like L~98-59 may seem unfavorable hosts to habitable planets, with the combination of many evolutionary factors, there is still a chance for them to host volatile-rich worlds, and they remain important targets to study terrestrial planets subject to frequent high levels of XUV radiation.

% Planet formation assumptions/justification
The presence of water on the L~98-59 planets is highly dependent on their composition, which is in turn dependent on their formation process and location.  There are three possible formation scenarios: 1) formation at their present location, 2) formation further out in the disk and inward migration to their current locations, and 3)  \added{gravitational scattering}\deleted{an instability in the disk} causing the planets to shift their orbits to their current locations \added{\citep{ogihara2009, chianglaughlin2013, ciesla2015}}. The L~98-59 planets orbit extremely close to their star, so it is most likely that they have migrated inward during the early stages of the system's formation \citep{ogihara2009, pulai2019}. Simulations by \citet{raymond2007} also show that it is difficult to form planets larger than 0.3 M$_\earth$ so close to a low-mass star. Given the predictions of studies by \citet{ogihara2009} and \citet{ciesla2015}, we consider there to be a significant likelihood for the L~98-59 planets to be composed of a non-negligible proportion of volatiles, including water, by the time they have fully formed and reached their current orbits.

% Limitations of this work (volatiles missing) + magma ocean potential
\subsection{Limitations of this work} \label{subsec:limits}

While our simulations show a range of outcomes that fall relatively in line with predictions from previous works, we note that they may not fully capture the specific details of the L~98-59 system's evolution. As established by \citet{barnesvpl}, the \texttt{AtmEsc} module of the \texttt{VPLanet} model is only an approximate description of atmospheric escape, and does not include some processes present in small, terrestrial planets, including the wavelength dependence of upper atmosphere heating and its variation with composition and atmospheric temperature structure, line cooling mechanisms, and other non-thermal escape processes. It also does not include CO$_2$, which is a very common volatile and likely to be found in the atmospheres of planets. But while the \texttt{VPLanet} model does not account for some known effects in the evolution of terrestrial planets, we believe that the results obtained give us a good estimate of the general state of the L~98-59 planets. More importantly, it provides us with a strong baseline for comparison with future observations of this system, or others like it. This general knowledge can 1) be a benchmark for what to expect when observing similar systems, 2) help us to interpret these future observations, and 3) refine our models and understandings of the evolution of small, multi-planet systems around M dwarfs.

Our simulations assume that no magma ocean is present on the planets. However like many planetary-sized bodies, the L~98-59 planets were likely molten right after their formation, a state known as a magma ocean \citep{solomon1979, wetherill1990, lammer2018}. This would affect their initial volatile budget, how much oxygen may accumulate, and how fast water is lost. During a magma ocean state, oxygen not only accumulates in the atmosphere but may also enter the melted surface, the latter which could outgas H$_2$O as a result (see \citet{barth2021}). These potential composition differences would affect the quantity of oxygen and water in the atmosphere, the quantity of oxygen stored in the planet, and the quantity of water produced by the magma ocean. Perhaps this would decrease the amount of oxygen in the atmosphere, and allow any water present to be ``protected'' by the magma ocean and retained longer. We note that future works could include this magma ocean state for a more complete picture of the L~98-59 planets' evolution.

\added{The age of L~98-59 is poorly constrained, and yet stellar age strongly affects what state the system's planets find themselves in. While the actual age of the star does not affect our results, since our simulations run from the star's PMS until the maximum age that the star could be, the age \textit{does} affect the implications of our results and the coupled interpretation of observations of these planets. Future JWST observations of the system would inform us on the age of the system and the starting water budget of the planets. Similarly, further constraints on the age of L~98-59 would affect our interpretation of our model results, thus further constraining the atmospheric state of the planets. For example, recent work by \citet{engle2023} finds an updated stellar age of around 4.94 Gyr for L~98-59, indicating that the system's planets could have experienced atmospheric escape for much longer than an age of $>$1 Gyr predicts. In the meantime, while our results provide a broad view of the potential evolution of the L~98-59 planets, this broad view will be further constrained when combined with the next observations of the system.}

Finally, in their analysis of a recent planet b transmission spectrum, \citet{Damiano2022} describe other potential models (besides cloud-free, H$_2$-dominated, and water-dominated) that include volatiles such as HCN, CO$_2$, CO, N$_2$, and CH$_4$, which our own simulations do not take into account. So while our results point to the possibility of a water vapor and oxygen-dominated atmosphere on L~98-59 b, our predictions may be limited because our model assumes a high initial water content on the planet to achieve this result and does not account for a varied set of volatiles.

%%%%%%%%%%%%%%%%
 % CONCLUSION
%%%%%%%%%%%%%%%%
\section{Conclusion} \label{sec:conclu}

In this paper, we model the evolution of the L~98-59 system to study the effects of atmospheric escape on the potential volatiles present on the inner three planets. Consisting of three small, transiting, and likely rocky planets (and a fourth non-transiting planet) orbiting their M-dwarf star, this system is an ideal candidate to better characterize the atmospheric evolution of multiple terrestrial planets within the same stellar environment.  

We run simulations with a parameter sweep of the system's characteristics, giving the planets an initial water budget of either 1, 10, or 100 TO each and evolving them over several billion years. We gather the following results:

% List out main results
\begin{enumerate}
    \item \added{All three planets have experienced}\deleted{All three planets are in a runaway greenhouse scenario from} strong atmospheric escape\deleted{,} due to their proximity to L~98-59 and the exposure to their star's XUV and flare activity.
    \item The smallest and closest planet, b, may counter-intuitively retain water longer than c and d, because its exposure to a higher XUV flux leads to increased radiative cooling, and thus a lower escape efficiency.
    \item Given enough starting water (100 TO) and depending on the system's age, all three planets may still have water leftover today, with the  highest chances of detection being for planet b.
    \item All three planets can accumulate significant quantities of oxygen created by water photolysis, from \added{15}\deleted{20} bars up to thousands of bars.
\end{enumerate}

Multi-planet systems can provide us with important insight on planet formation and evolution, orbital dynamics, and planetary architectures. Moreover, planets around bright, nearby stars such as L~98-59 are ideal targets for atmospheric characterization with emission and transmission spectroscopy \citep{kostov2019b}. This system serves as an excellent case to study the development of multiple terrestrial planets and their atmospheres. 

Further, the presence of an atmosphere on any of the L~98-59 planets is highly dependent on the way the planets evolved in the presence of their star's stellar activity, and specifically, the initial composition with which they began evolving. Therefore, our work studying the effects of flares and strong XUV radiation on volatile loss or accumulation for these planets may help us to better constrain certain characteristics of the system, such as stellar age, initial planetary composition, and planetary formation processes, in future observations with JWST.

%%%%%%%%%%%%%%%%
 % ACKNOWLEDGEMENTS
%%%%%%%%%%%%%%%%
\section*{Acknowledgements}

The material is based upon work supported by NASA under award number 80GSFC21M0002. \added{This research has made use of the NASA Exoplanet Archive, which is operated by the California Institute of Technology, under contract with the National Aeronautics and Space Administration under the Exoplanet Exploration Program.} This research was carried out in part at the Jet Propulsion Laboratory, California Institute of Technology, under a contract with the National Aeronautics and Space Administration (80NM0018D0004). RB acknowledges support from NASA grant 80NSSC23K0261. L.N.R.A. acknowledge the support of UNAM DGAPA PAPIIT project IN110420. L.N.R.A. thanks CONACYT’s graduate scholarship program for its support.
E.F.F thanks Thaddeus D. Komacek for insightful discussions regarding simulation results\added{, and Siddhant Solanki for their help with optimizing our code and running a set of our simulations during the revisions of this paper. We thank the anonymous referee for their insightful comments and valuable suggestions, which improved the overall quality of this paper}. 

\textit{Software:} VPLanet \citep{barnesvpl}, mercury \citep{chambers2000}, jupyter \citep{Kluyver2016}, NumPy \citep{harris2020array}, SciPy \citep{SciPy-NMeth2020}, Matplotlib \citep{Hunter2007}, Pandas \citep{reback2020pandas}, SymPy \citep{sympy}, seaborn \citep{Waskom2021}, fitter \citep{thomas_cokelaer_fitter}.

\bibliography{references}{}
\bibliographystyle{aasjournal}

%------------------------------------------------

%% This command is needed to show the entire author+affiliation list when
%% the collaboration and author truncation commands are used.  It has to
%% go at the end of the manuscript.
%\allauthors

%% Include this line if you are using the \added, \replaced, \deleted
%% commands to see a summary list of all changes at the end of the article.
%\listofchanges

\end{document}